\documentclass[apj]{emulateapj}

\usepackage{apjfonts}

\usepackage{natbib}

\shorttitle{Optical Identification of PSR J1544+4937}
\shortauthors{Tang et al.}

\begin{document}

\title{Identification of the Optical Counterpart of Fermi Black Widow Millisecond Pulsar PSR J1544+4937}


\author{Sumin Tang\altaffilmark{1,2}, 
David L. Kaplan\altaffilmark{3}, Sterl Phinney\altaffilmark{1},  Thomas A. Prince\altaffilmark{1}, Rene P. Breton\altaffilmark{4}, 
Eric Bellm\altaffilmark{1}, \\
Lars Bildsten\altaffilmark{2, 5}, 
Yi Cao\altaffilmark{1}, 
A.~K.~H. Kong\altaffilmark{6}, 
Daniel A. Perley\altaffilmark{1,7}, 
Branimir Sesar\altaffilmark{8}, William M. Wolf\altaffilmark{5}, and T.-C. Yen\altaffilmark{6}}

\altaffiltext{1}{Division of Physics, Mathematics, \& Astronomy, California Institute of Technology, Pasadena, CA 91125, USA}
\altaffiltext{2}{Kavli Institute for Theoretical Physics, University of California, Santa Barbara, CA 93106, USA}
\altaffiltext{3}{Physics Department, University of Wisconsin--Milwaukee, Milwaukee, WI 53211, USA}
\altaffiltext{4}{School of Physics and Astronomy, University of Southampton, SO17 1BJ, UK}
\altaffiltext{5}{Department of Physics, University of California, Santa Barbara, CA 93106, USA}
\altaffiltext{6}{Institute of Astronomy and Department of Physics, National Tsing Hua University, Hsinchu 30013, Taiwan}
\altaffiltext{7}{Hubble Fellow}
\altaffiltext{8}{Max Planck Institute for Astronomy, K\"{o}nigstuhl 17, D-69117 Heidelberg, Germany}

\begin{abstract}
We report the optical identification of the companion to the {\it Fermi} black widow millisecond pulsar PSR J1544+4937.
We find a highly variable source on Keck LRIS images at the nominal pulsar position,
with 2 magnitude variations over orbital period in the B, g, R, and I bands.
The nearly achromatic light curves are difficult to explain with a simply irradiated hemisphere model, 
and suggest that the optical emission is dominated by a nearly isothermal hot patch on the surface of the companion facing the pulsar.
We roughly constrain the distance to PSR J1544+4937 to be between 2 and 5 kpc.
A more reliable distance measurement is needed in order to constrain the composition of the companion.
\end{abstract}


\section{Introduction}

Millisecond pulsars (MSPs) are believed to be old neutron stars spun-up by accreting matter from a companion star,
and thus are the end products of the accretion process observed in low-mass X-ray binaries \citep[LMXBs;][]{1982Natur.300..728A, 2009Sci...324.1411A}.
They are unique laboratories of high--energy plasma physics and magnetic reconnection. 
Only in MSPs do the emission regions of $\gamma$--rays probe the magnetic fields close to the neutron star.
The {\it Fermi} Large Area Telescope \citep[LAT;][]{2009ApJ...697.1071A} has revolutionized the study of MSPs, leading to the discovery of 43 new MSPs, 
and strikingly, 40\% of the newly identified {\it Fermi} MSPs with measured orbital parameters turned out to be 
``black widows'' (BWs; companion mass $M_{C}<0.1\ M_{\odot}$) or ``redbacks'' (RBs; $M_{C}\approx0.1-0.4\ M_{\odot}$), 
in which the pulsar radio emission is eclipsed by a wind from the companion 
\citep{2012arXiv1205.3089R, 2013IAUS..291..127R}.

PSR J1544+4937 is a {\it Fermi} LAT $\gamma$--ray source,
identified as a radio millisecond pulsar (MSP)
with a 2.16-ms spin period by the Giant Metrewave Radio Telescope \citep{2013ApJ...773L..12B}.
The radio timing observations give an orbital period of 2.9 hrs with an eclipse duration of 22 min (13\% of its orbit), 
and a minimum companion mass of 0.017 $M_\odot$ ($i=90^\circ$ and $M_{NS}=1.4\ M_\odot$),
making it a BW pulsar. 
No optical counterpart was detected in the archival SDSS images to 
limiting magnitudes of $g'=22.5$, $r'=22.0$ and $i'=21.5$ \citep{2013ApJ...773L..12B}. 

We have obtained deep optical images to identify the optical counterpart of PSR J1544+4937,
and to constrain the properties of the system.
A highly variable source is revealed in our optical images at the radio pulsar position.
We describe the optical observations and photometry in \S 2.
We discuss the distance to the source and present the light curve modeling in \S3,
and the mass-radius relation of the companion in \S4.
A final discussion of the results is in \S5.

\section{Optical Photometry}

We observed PSR J1544+4937 using 
the Low Resolution Imaging Spectrometer \citep[LRIS;][]{1995PASP..107..375O} on the Keck I 10--m telescope on 4 nights in 2013:
May 9, Aug 4, Sep 4, and Sep 9.
We used the {\it B}, {\it g}, {\it R}, and {\it I} filters, with exposure times ranging from 190\,s to 300\,s.
All the {\it I} band images were taken simultaneously with {\it g} band images,
while all the {\it R} band images were taken simultaneously with {\it B} band images,
and therefore the measured $g-I$ and $B-R$ colors are not affected by variability.
The {\it g} band images taken on May 9 were not accompanied by {\it I} band images due to filter wheel problems that night.
We also took 30\,s exposures of the same fields in order to get a large sample of unsaturated stars for zero-point calibration.
The images were taken at airmass\,$<2$ with seeings of 0.6$\arcsec$--0.8$\arcsec$.

\begin{figure*}
\epsscale{1.2}
\plotone{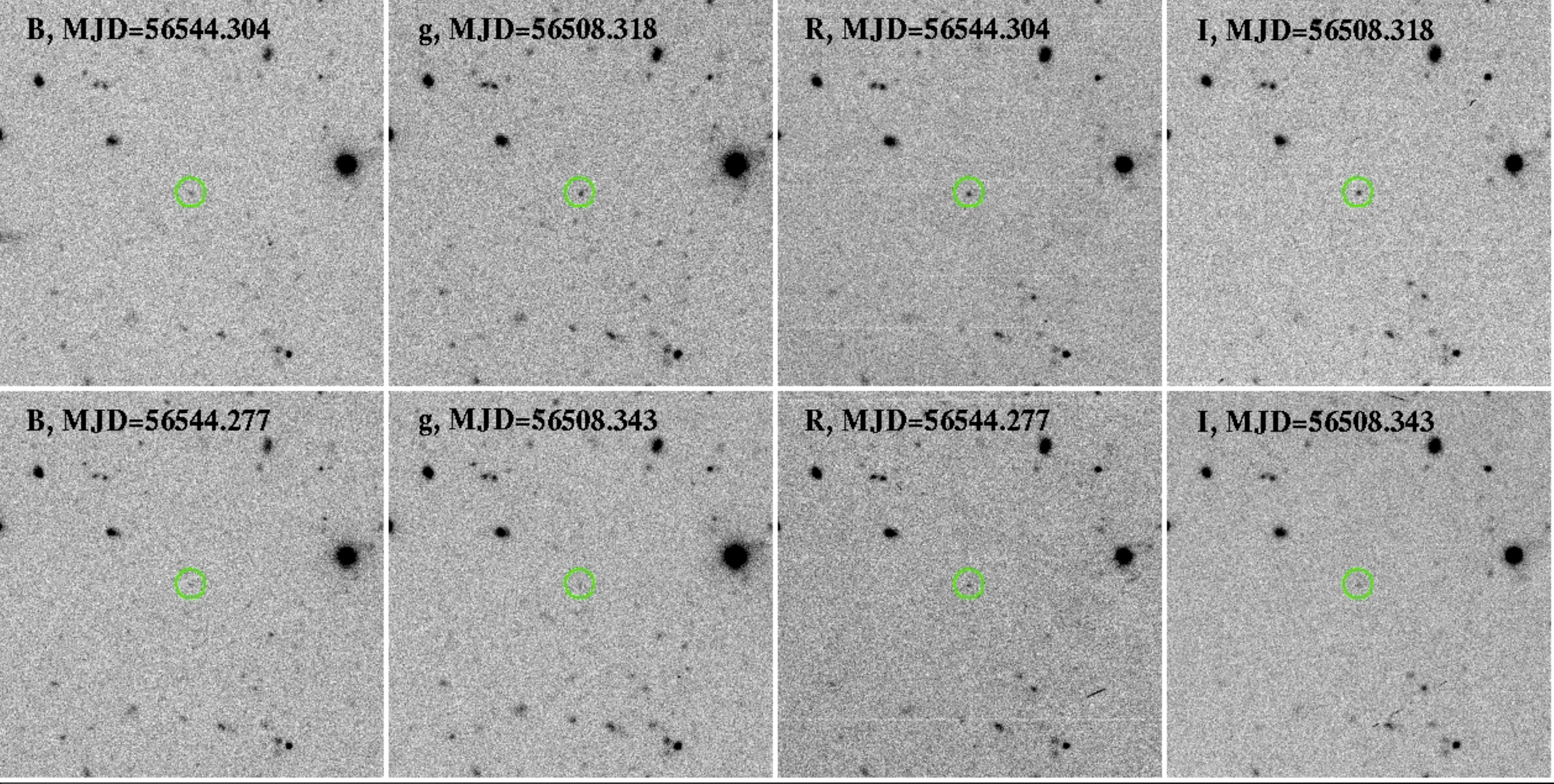}
\caption{Keck LRIS images of PSR J1544+4937, revealing a variable optical source at the nominal pulsar position.
From the left to the right: B, g, R, and I band images, respectively.
In each band, a pair of images taken within 1 hour are shown, with the brighter one on the top.
Green circles denote the radio position of the pulsar from  \citet{2013ApJ...773L..12B} with radii of 3\arcsec.
Each image is 80\arcsec$\times$80\arcsec.
\label{fig1}}
\end{figure*}

The data were reduced via standard techniques using  LPip\footnote{http://www.astro.caltech.edu/~dperley/programs/lpipe.html}. 
Photometry was performed first on the 30\,s exposure images in IDL using a 1\arcsec\ radius aperture relative to unsaturated SDSS stars in the field of view, 
with {\it gri} transformed to {\it BRI} using the Lupton transformation equations\footnote{http://www.sdss.org/dr4/algorithms/sdssUBVRITransform.html\#Lupton2005},
and then performed on the 190\,s to 300\,s exposure images relative to unsaturated stars in the 30\,s exposure images.

As shown in Figure 1, a highly variable source is detected within 0.1\arcsec\ of the radio position of the pulsar \citep[$\alpha = 15^{h} 44^{m} 04.487^{s}$, $\delta = 49^{\circ} 37\arcmin 55\farcs25$;][]{2013ApJ...773L..12B} in all four bands,
consistent with our typical astrometric uncertainty.
The counterpart profiles are consistent with a point source, with no detectable position drift within 0.1\arcsec.
In Figure 2, we show the LRIS optical light curves, folded using the radio timing ephemeris \citep{2013ApJ...773L..12B}
and shifted by 0.25 so that phase $0$ corresponds to the pulsar superior conjunction (radio eclipse of the pulsar by the companion).
As shown in the middle and bottom panels of Figure 2, the light curves are achromatic within photometric uncertainties over the orbital period,
with $g-I =1.91\pm0.08$\,mag and $B-R =1.65\pm0.13$\,mag,
where the quoted errors are the standard deviations of the mean.
To test the reliability of our photometry, especially at the faint end, 
we performed the same photometric procedures on several neighbor stars with comparable magnitudes of the targeted source,
and all of them are indeed constant within photometric errors.

We also obtained one $r'$-band image with an 860\,s exposure 
using the MegaCam on the Canada-France-Hawaii Telescope \citep[CFHT;][]{2003SPIE.4841...72B} 
on 2013 Feb 11. 
A point source is detected at the nominal radio position with $r'=23.86\pm0.10$ mag,
which corresponds to $R=23.59\pm0.10$\,mag using the Lupton transformation equations.
The converted R band magnitude from CFHT is shown as a red square in Figure 2.

\begin{figure*}
\epsscale{1.}
\plotone{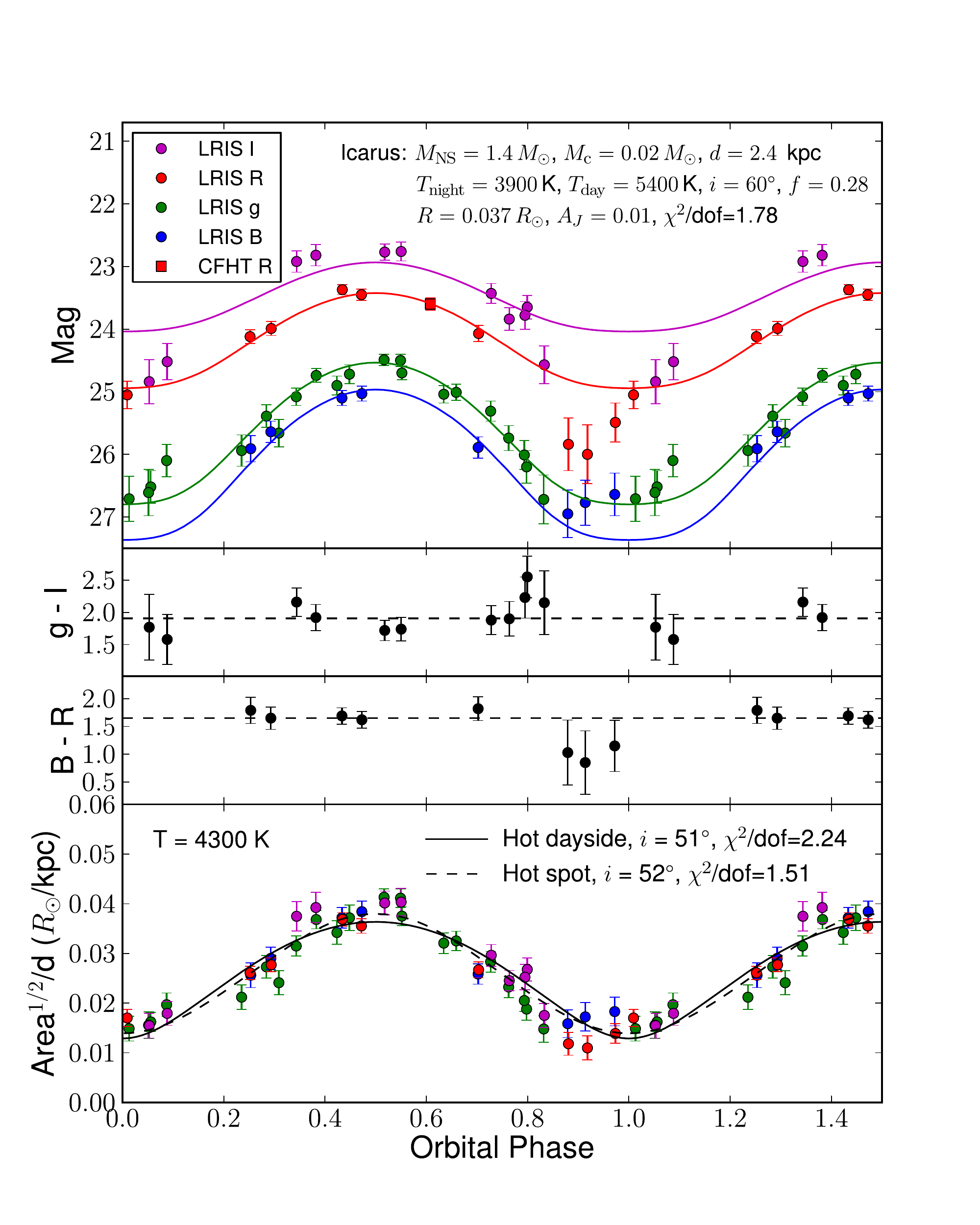}
\caption{{\it Top:} Folded optical light curves of PSR J1544+4937. 
The I, R, g, and B data shown in solid circles were obtained with Keck LRIS; 
the one $r'$-band point shown in red square was obtained with CFHT MegaCam.
A best--fit model from Icarus with $\chi^2/dof=1.78$ ($dof=42$) is shown in solid lines, with model parameters listed.
{\it Middle:} simultaneous measured LRIS $g - I$ color and $B-R$ color vs. orbital phase, respectively.
The weighted mean of $g - I = 1.91$ and $B - R = 1.65$ are shown in dashed lines.
The colors are consistent with no evolution.
{\it Bottom:} The ratio of the square root of the projected area over distance in units of $R_\odot$/kpc as a function of orbital phase, assuming all the optical emission is from regions with $T = 4300$\,K.
The BTSettl atmosphere model is used in the conversion \citep{2011ASPC..448...91A}.
Symbols are the same as in the top panel.
The solid line shows the best-fit hot dayside model (an isothermal dayside with $T = 4300$\,K and negligible emission from the night side) with inclination angle $i=51^\circ$ and $\chi^2/dof=2.24$ ($dof=48$).
The dashed line shows a best-fit hot spot model (an isothermal hot spot with $T = 4300$\,K  facing the neutron star, and negligible emission from other regions),
with a radius of 57$^\circ$, centered at latitude 38$^\circ$,  and viewed at $i=52^\circ$; The corresponding reduced chi-squared is $\chi^2/dof=1.51$ ($dof=46$).
\label{fig2}}
\end{figure*}

\section{Light Curve Modeling}

\subsection{Distance to PSR J1544+4937}
PSR J1544+4937 is located at Galactic coordinates $l=79^\circ$ and $b=50^\circ$,
with an observed dispersion measure (DM) of PSR J1544+4937 is 23.23 cm$^{-3}$ pc  \citep{2013ApJ...773L..12B}.
The thicker layer of free electrons in the Milky Way, the so-called warm ionized medium,
has a scale-height of about 1.8 kpc \citep{2008PASA...25..184G}. 
At vertical distances out of the galactic plane larger than 1--2 kpc, 
the DM vs distance curve flattens \citep{2008PASA...25..184G},
making DM an insensitive indicator of distance. 
PSR J1544+4937 unfortunately falls into this insensitive region.
The NE2001 model by \citet{2002astro.ph..7156C} gives a distance estimate of $1.2^{+0.4}_{-0.3}$\, pc,
while \citet{2008PASA...25..184G} suggests that the distance is 2--5 kpc (see Figure 1 in their paper for DM$\sin|b|=17.84$ cm$^{-3}$ pc at $|b|>40^\circ$).
For comparison, B1508+55 ($l=91^\circ$ and $b=52^\circ$), 
the pulsar with a reliable parallax nearest to PSR J1544+4937 ($8^{\circ}$ away), 
has a DM$=$19.61 cm$^{-3}$ pc, and a parallax distance of $2.10^{+0.13}_{-0.14}$ kpc \citep{2009ApJ...698..250C}.
This suggests that J1544+4937, which has a higher DM than B1508+55, is likely to have a distance larger than 2 kpc.
Therefore, we adopt a distance of 2--5\,kpc for PSR J1544+493.

\subsection{Effective Temperature and Heating Efficiency}

PSR J1544+4937 is located at high galactic latitude ($b=50^\circ$), 
and the extinction correction in this direction is negligible with $E(B-V)=0.015$\,mag \citep{2011ApJ...737..103S}.
The measured $g-I$ and $B-R$ colors imply a nearly constant effective temperature of $T_{eff}=4300\pm200$\,K over different orbital phases.
If the temperature is dominated by isotropic pulsar heating, 
given the radio timing results from \citet{2013ApJ...773L..12B},
the inferred heating efficiency is $\eta=4\pi\sigma a^2 T_{irr}^4/\dot{E}=0.14\pm0.03$,
where $a$ is the orbital separation, $T_{irr}$ is the irradiation temperature ($T_{irr}^4 = T_{day}^4-T_{night}^4$), and $\dot{E}$ is the spin down luminosity.
Such an efficiency is comparable to the values estimated in other BW systems \citep{2013ApJ...769..108B}.


\subsection{Icarus Modeling}

We first tried to fit these light curves using Icarus, a newly
developed binary light curve synthesis code \citep{2012ApJ...748..115B}.
Only Keck LRIS data were used in the modeling.
Pulsar heating is taken into account with a heating efficiency $\eta$ and 
a $\cos\theta$ factor to account for the projected illuminated area,
where $\theta$ is the angle between the surface normal and the direction to the pulsar.
BTSettl atmosphere models are used \citep{2011ASPC..448...91A}.
Heat redistribution is not included.
There are seven free parameters in the model:
the mass ratio $q$,
the orbital inclination $i$,
the distance $d$,
the companion's filling factor $f=R/R_L$ relative to its Roche lobe radius $R_L$,
its dayside and nightside temperatures $T_{day}$ and $T_{night}$,
and the reddening in J band $A_J$ to the system assuming $A_J/A_V=0.282$.
Given the high galactic latitude and low extinction ($E(B-V)=0.015$\,mag) in this direction, we restrict $A_J<0.1$.
Given the eclipses observed in radio, the binary cannot be face-on, and thus we assume $\cos(i)<0.8$ ($i>37^\circ$).
We also assume that the companion is tidally locked.

\begin{figure*}
\epsscale{0.7}
\plotone{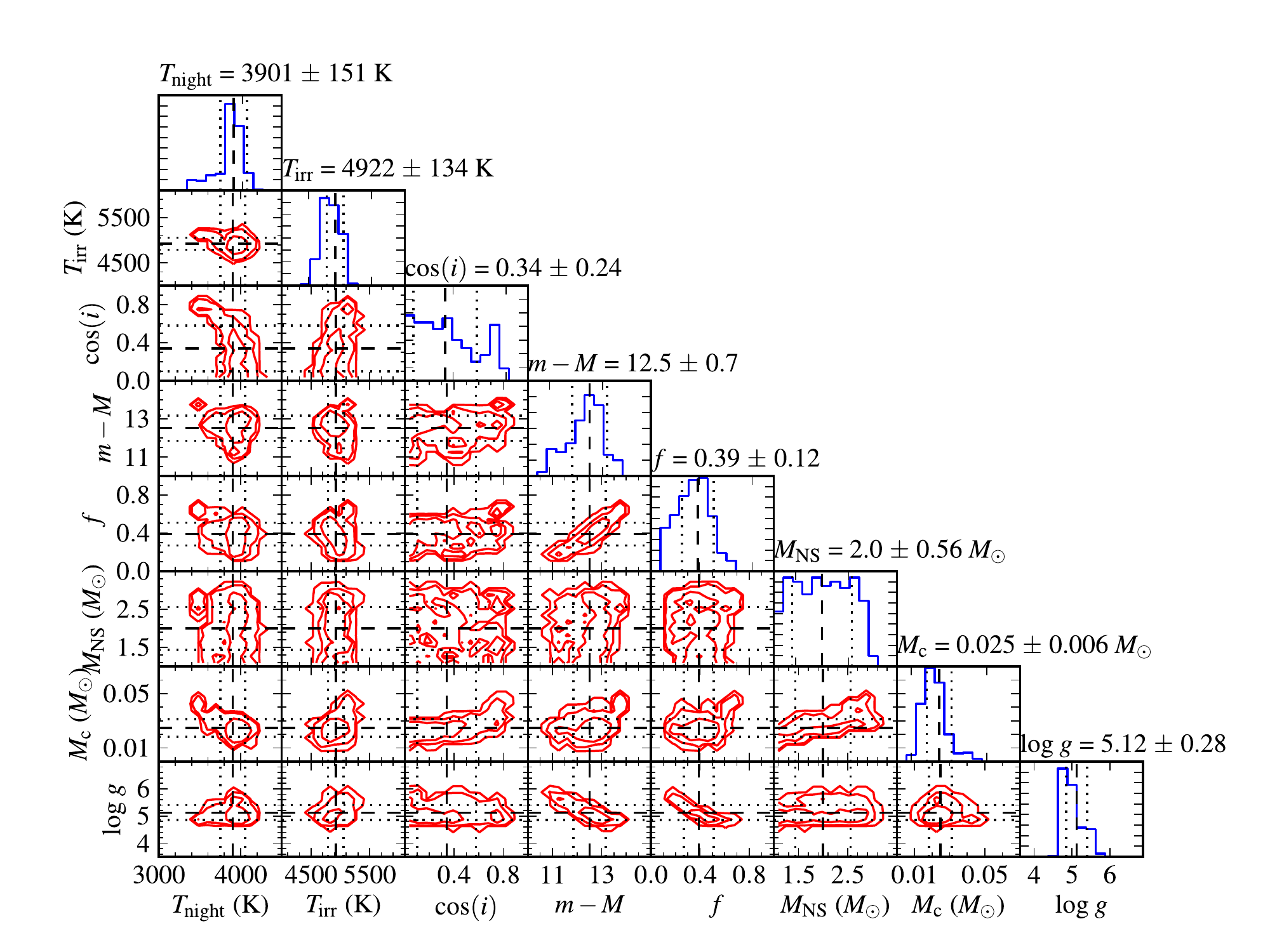}
\caption{One- and two-dimensional distributions of the key parameter values resulting from our Icarus MCMC fitting.
Values of the mean and standard deviation of each parameter are given.
Red contours denote the 1$\sigma$, 2$\sigma$ and 3$\sigma$ confidence levels.
In all of the plots, the black dashed lines show the means and the black dotted lines the $\pm1\sigma$ limits.
\label{fig3}}
\end{figure*} 

There is no set of parameters in Icarus with $i>37^\circ$ that fits the light curve well, and even the best fits have $\chi^2/{\rm dof} = 1.78$ (dof$=42$).
Due to the assumed temperature distribution, all Icarus models predict larger amplitude in bluer bands.
One best--fit example by Icarus is shown in the top panel of Figure 2.
The resulting one- and two-dimensional marginalized confidence contours of the parameters are shown in Figure 3.
Given the dayside temperature distribution assumed, the irradiation temperature $T_{irr}$ 
is mostly affected by the observed color and therefore is relatively well constrained.
There is no constraint on the inclination, distance (which is degenerate with the filling factor), 
and neutron star mass (which is degenerate with the companion mass).

If we remove the constraints on inclination angles in Icarus, we could get a group of relatively better fits to the light curves ($\chi^2/{\rm dof} = 1.5-1.6$) 
with cool dayside ($T_{night}<2000$\,K) and low inclination angles ($i=15-30^\circ$).
Such a set of parameters leads to negligible flux from the night side, therefore relatively more achromatic light curves.
However, given the observed eclipses in radio \citep{2013ApJ...773L..12B}, this is unlikely to be true.

\subsection{Isothermal Patch(es) on the Companion}

Alternatively, the achromatic light curves can be explained in models with hot patches at $T\approx4300$\,K  on the surface of the companion, 
and negligible radiation from other regions. 
In such scenarios, the optical variability only depends on the projected area of the hot patches along line of sight.
We can then derive the projected solid angle (effective area over distance square, $A_{proj}/d^2$) of these hot patches as a function of orbital phase from the optical light curves.
In the bottom panel of Figure 2, 
we show the resulted ratio of the square root of the projected area over distance ($A_{proj}^{1/2}/d$) in units of $R_\odot$/kpc.
We assumed $T=4300$\,K and the BTSettl atmosphere model \citep{2011ASPC..448...91A}.

We tried two toy models of hot patches to fit the $A_{proj}^{1/2}/d$ curve. 
First, we tried a hot dayside model with an isothermal temperature profile at $T = 4300$\,K and negligible emission from the night side.
Such a hot dayside could be caused by effective heat transportation/redistribution on the dayside.
A spherical star was used.
We obtained the best-fit at inclination angle $i=51^\circ$ with $\chi^2/dof=2.24$ ($dof=48$), as shown in the solid line in the bottom panel of Figure 2.
We noticed that the optical light curves might be slightly asymmetric,
with an optical minimum earlier than the radio eclipse.
Future deep imaging observations near the minimum will allow one to investigate the possible asymmetry.

Second, we tried a hot spot model with an isothermal profile at $T = 4300$\,K  facing the neutron star, and negligible emission from other regions.
Such a hot spot could be caused by the magnetic field of a companion, 
where the magnetic pole(s) are potentially heated by the magnetically confined pulsar wind.
We treated the hot spot radius ($r_{spot}$), central latitude ($l_{spot}$), and inclination angle $i$ as free parameters.
We obtained the best-fit at $i=52^\circ$,  $r_{spot}=57^\circ$, and $l_{spot}=38^\circ$,  with $\chi^2/dof=1.51$ ($dof=46$),
as shown in the dashed line in the bottom panel of Figure 2.

A third option for the hot patch is a stripe irradiated by the energetic, narrow pulsar beam rather than the hypothetically more uniform pulsar wind. 
Such a scenario has been demonstrated in the case of the redback PSR~J1740$-$5340, 
where the authors conclude that there is a very  narrow, but longitudinally extended stripe covering $\lesssim 1\%$ of the companion surface \citep{2003ApJ...589L..41S}.
A similar phenomenon could be acting in PSR~J1544$+$4937. 
Due to relatively large scattering in the data (as shown in the bottom panel in Figure 2), 
it is difficult to distinguish different models, and therefore we did not attempt to model the stripe.

\section{Mass-Radius Relation of the Companion}

The lowest possible mean density of the companion is reached when the companion fills its Roche lobe.
This minimum mean density can be approximately set by the orbital period \citep{1972ApJ...175L..79F}: 
$\bar{\rho}_{c, min}=3\pi/(0.462^3GP_b^2)=13.1$ g cm$^{-3}$,
as shown in the black solid line in Figure 4,
which puts an upper limit for possible companion radius at a given companion mass.

\begin{figure*}
\epsscale{1.2}
\plotone{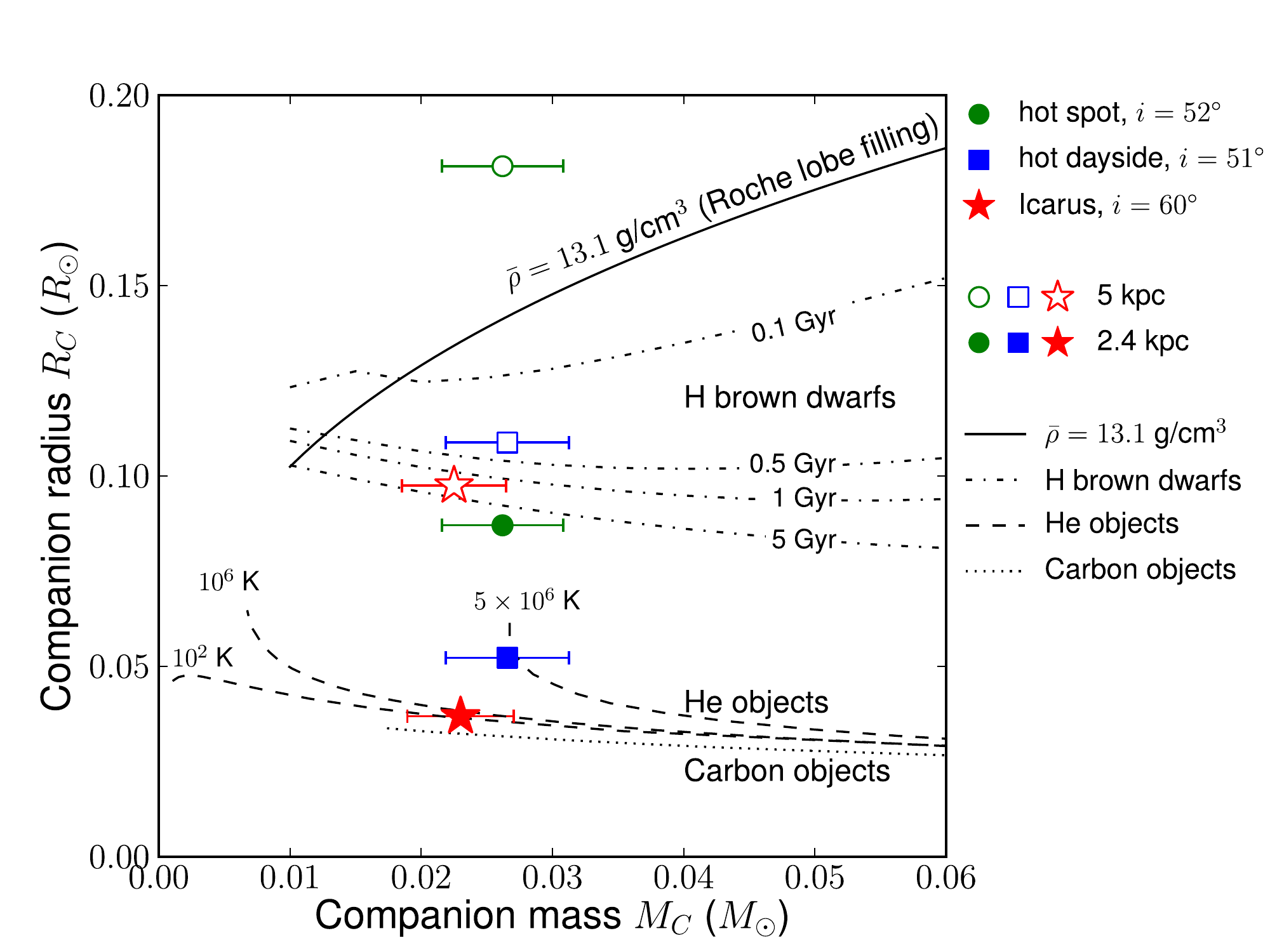}
\caption{Mass--radius relation for the companion of PSR J1544+4937 derived from different light curve modelings. 
The green circles are from the best hot spot model with $T_{spot}=4300$\,K,  radius of 57$^\circ$, centered at latitude 38$^\circ$ facing the neutron star,  and viewed at $i=52^\circ$.
The blue squasars are from the best-fit hot dayside model with $T_{day}=4300$\,K and viewed at $i=51^\circ$.
The red stars are from Icarus with $60^\circ$.
Open symbols, and solid symbols, denote models with distances of 5 kpc, and 2.4 kpc, respectively.
The errorbars are derived from assumed neutron star masses of $M_{NS}=1.7\pm0.3\ M_\odot$.
The black solid line show a mean density of $\bar{\rho}=13.1$ g cm$^{-3}$, which is approximately the mean density of a Roche-lobe filling star at an orbital period of $P=2.9$ h,
and therefore is the lower limit of the companion of PSR J1544+4937.
The other black curves show the theoretical mass--radius relations for brown dwarfs, Helium and carbon objects.
The dash-dotted lines are brown dwarfs with ages 0.1, 0.5, 1.0 and 5.0 Gyr, from top to bottom,
from MESA \citep{2011ApJS..192....3P, 2013ApJS..208....4P};
the results are consistent with \citet{2000ApJ...542..464C} with extended coverage at lower masses.
Dashed lines show Helium objects with core temperatures of $5\times10^6$, $10^6$, and $10^2$ K, from top to bottom, 
and the dotted line show carbon objects with core temperatures of $10^6$ \citep{2003ApJ...598.1217D}.  
\label{fig4}}
\end{figure*}

Given the radio timing constraints, the mass of the companion depends on the neutron star mass and binary orbital inclination.
The radius of the companion can be estimated from optical light curves, 
and are most sensitive to the distance, inclination and the assumed optical emitting model.
To simplify the case, we assumed a neutron star mass of $M_{NS}=1.7\pm0.3\ M_\odot$,
and two possible distances of $d=2.4$ kpc and 5 kpc.
We adopted the best-fit hot spot and hot dayside models discussed in \S 3.4,
and an fixed inclination angle of $i=60^{\circ}$ for Icarus.
The resulted companion masses and radii are shown in Figure 4,
with green circles, blue squares, and red stars
for hot spot, hot dayside, and Icarus models, respectively.
Open symbols are for $d=5$ kpc, and solid symbols are for $d=2.4$ kpc.

For comparison, we also plotted the theoretical mass-radius relations for 
hydrogen brown dwarfs from the Modules for Experiments in Astrophysics \citep[MESA;][]{2011ApJS..192....3P, 2013ApJS..208....4P}, 
and helium and carbon objects from \citet{2003ApJ...598.1217D}.
Note that these theoretical curves were derived for isolated objects,
and pulsar heating could inflate the companion.

As shown in Figure 4, if the optical emission is from a heated dayside (hot dayside or Icarus models),
the companion is likely a hydrogen brown dwarf if $d\approx5$ kpc,
or a helium or carbon object if $d\approx2-3$ kpc.
If the optical emission is dominately from a small hot spot,
at a given distance,
the inferred radii are much larger than the radii given by the hot dayside or Icarus models, and could be a hydrogen brown dwarf at $d\approx$2.4\,kpc.
A hydrogen companion could be the descendent of an accreting MSP LMXB such as SAX J1808.4--3658 \citep{2001ApJ...557..292B}.
On the other hand, a helium or carbon companion could be the descendent of 
an ultra compact X-ray binary ($P\approx0.5-0.9$ days),
where the donor had substantial mass loss via Roche lobe overflow and pulsar wind evaporating as it evolved
\citep{2012ApJ...753L..33B, 2013MNRAS.433L..11B, 2013ApJ...775...27C}.

\section{Discussion}

We identified the optical counterpart of PSR J1544+4937 and obtained multi-band optical light curves,
which show 2\,mag modulations in the B, g, R and I bands at the 2.9 hr orbital period given by the radio timing ephemeris (Figure 2).  
The optical maximum occurs around the inferior conjunction, where the pulsar is located between the companion and the Earth,
suggesting that the optical variation is due to pulsar heating on the dayside of the companion facing the pulsar.

An intriguing feature of this system revealed by optical photometry is that
there is no significant color variation over the orbital period.
The observed color is consistent with an effective temperature of $\approx$4300\,K. 
The standard pulsar heating models with $T_{irr} \propto \cos\theta$ on the dayside do not work well for the observed light curves. 
A possible model is that the optical radiation is dominated by a heated magnetic pole of the companion (dashed line in the bottom panel of Figure 2). 
Assuming the radio eclipses are caused by cyclotron/synchrotron absorption, 
\citet{2013ApJ...773L..12B} estimate a required magnetic field of $B\sim11$ G in the vicinity of the companion at a scale of 0.5 $R_\odot$.
If the required magnetic field is provided by the magnetosphere of the companion  \citep{1994ApJ...422..304T},
the magnetic field near the surface of the companion ($\sim0.05-0.1\ R_\odot$; see \S 4) 
is therefore 25--100 times higher, i.e. $B\sim$0.3--1.1\, kG.
This is comparable to magnetic fields in brown dwarfs \citep{2010AA...522A..13R}. 

Besides, it appears the companion is unlikely to be Roche-lobe filling, 
unless the distance is larger than 5 kpc or its optical radiation is dominantly from a small region on the surface of the companion,
as shown in Figure 4.
Previous study has found that not all BWs and RBs fill their Roche lobes \citep{2013ApJ...769..108B}. 
As discussed by \citet{2013ApJ...769..108B}, 
this has direct implications for the radio eclipses, 
as the ionized gas cloud floating around cannot originate from loosely bound matter at the surface of a Roche-lobe filling star,
and has to have been pulled off the surface or been generated by a wind from the companion.


Only a few BWs have optical light curves with good color coverage, with the best one being PSR J1311--3430. 
Interestingly, the standard pulsar-heating model cannot fit the optical light curves of PSR J1311--3430 either \citep{2012ApJ...760L..36R}.
One model the authors adopted is to add a cold spot near L1.
This led to reduced temperature differences on the dayside,
which is similar to our toy models of isothermal hot dayside or spot models. 
Optical imaging of BWs in the future with better color coverage can enable more detailed study of the temperature distribution of the companion,
which will help understand the pulsar heating process, and heat reprocessing/transportation on the companion.

\acknowledgments

We thank the referee for helpful comments.
ST acknowledges support by the NASA grant NNX12AO76G.
DLK is supported by the National Science Foundation grant AST-1312822. 
AKHK and TCY are supported by the Ministry of Science and Technology
of the Republic of China (Taiwan) through grants
100-2628-M-007-002-MY3, 100-2923-M-007-001-MY3, and
101-2119-M-008-007-MY3.
DAP is supported by NASA through Hubble Fellowship grant HST-HF-51296.01-A.
We are grateful to S. R. Kulkarni for his help on obtaining Keck data.
The W. M. Keck Observatory is operated as a scientific partnership among the California Institute of Technology, the University of California, and NASA; the Observatory was made possible by the generous financial support of the W. M. Keck Foundation.

{\it Facilities:}  \facility{Keck (LRIS)}, \facility{CFHT (MegaCam)}.

\end{document}